\documentstyle[epsfig,letters]{mn1}

\title[Intrinsic Shapes of Galaxy Clusters]{Galaxy Clusters: Oblate or Prolate?}
\author[Cooray]{Asantha R. Cooray\\
Department of Astronomy and Astrophysics, University of
Chicago, Chicago IL 60637, USA. E-mail: asante@hyde.uchicago.edu}
\date{Accepted for Publication in {\it Monthly Notices of Royal
Astronomical Society}}
\begin{document}
\maketitle

\begin{abstract} 

It is now well known that a combined analysis of 
the Sunyaev-Zel'dovich (SZ) effect and the X-ray emission observations
can be used to determine the angular diameter distance to galaxy
clusters, from which the Hubble constant is derived. 
Given that the SZ/X-ray Hubble constant is determined through 
a geometrical description of clusters, the accuracy to
which such distance measurements can be made depends on
how well one can describe intrinsic cluster shapes. 
Using the observed X-ray isophotal axial ratio distribution for a
sample of galaxy clusters, we discuss intrinsic cluster shapes and, 
in particular, if clusters can be described by axisymmetric 
models, such as oblate and prolate ellipsoids. These models are 
currently favored when determining the SZ/X-ray 
Hubble constant. We show that the current observational data on the 
asphericity of galaxy clusters suggest that clusters are more consistent with a
prolate rather than an oblate distribution.
We address the possibility that clusters are intrinsically triaxial by viewing
triaxial ellipsoids at random angles with the intrinsic axial ratios 
following an isotropic Gaussian distribution. 
We discuss implications of our results on current attempts at
measuring the Hubble constant using galaxy clusters
and suggest that an unbiased estimate of the Hubble
constant, not fundamentally limited by projection effects,
would eventually be possible with the SZ/X-ray method.

\end{abstract}


\section{Introduction}

Over the last few years, there has been a tremendous increase in 
the study of galaxy clusters as cosmological probes, initially through the
use of X-ray emission observations, and in recent years,
through the use of Sunyaev-Zel'dovich (SZ) effect.
Briefly, the SZ effect is a distortion of the cosmic microwave background
(CMB) radiation by inverse-Compton scattering of 
thermal electrons within the hot intracluster medium
(Sunyaev \& Zel'dovich 1980; see Birkinshaw 1998 for a recent review). 
The initial motivation for the study of SZ effect was to establish a
cosmic origin to the cosmic microwave background (CMB), rather
than a galactic one. It was later realized, however,
 that by combining the SZ intensity change and the X-ray
emission observations, and solving for the number density distribution
of electrons responsible for both these effects after 
assuming a certain geometrical shape, 
angular diameter distance, $D_{\rm A}$, to galaxy clusters 
can be derived (e.g., Cavaliere {\it et al.} 1977; Silk \& White
1978; Gunn 1978). Combining the distance measurement with redshift
allows a determination of the Hubble constant, $H_0$, through the
well known angular diameter distance relationship with redshift, 
and after assuming a geometrical world model with values for the 
cosmic matter density, $\Omega_m$, and the cosmological constant, 
$\Omega_\Lambda$. On the other hand, angular diameter distances with 
redshift for a sample of clusters,
over a wide range in redshift, can be used to constrain
cosmological world models; An approach essentially similar to the
one taken by two groups to constrain $\Omega_m$ and $\Omega_\Lambda$ using
luminosity distance relationship of Type Ia supernovae as a function
of redshift (Perlmuetter et al. 1998; Riess et al. 1998).

The cosmological parameter measurements using Type Ia supernovae 
are based on the fact that these supernovae are standard candles,
or standard candles after making appropriate corrections (see,
Branch 1999 for a recent review). 
Since the SZ/X-ray distance measurements are based on geometrical
method, one requires detailed knowledge on galaxy cluster shapes.
However, such details are not always available; in some cases, e.g.,
the cluster inclination angle, such details are not likely to be ever 
available. Also, given that the two effects involved are due to the spatial
distribution of electrons and their thermal structure,
additional details on the physical properties of electron
distribution are needed. Thus, the accuracy to which the Hubble constant
can be determined from the SZ/X-ray route
depends on the assumptions made with regards to the cluster shape
and its physical properties, or how well such information can
be derived a priori from data. 
Current measurements on the Hubble constant using cluster X-ray
emission and SZ are mostly based on the assumption of an
isothermal temperature distribution and a spherical geometry for
galaxy clusters. In recent years, improvements
to the spherical assumption have appeared in the form of axisymmetric 
elliptical models (e.g., Hughes \& Birkinshaw 1998).

Using analytical and numerical tools, several investigations 
have now studied the accuracy to which the Hubble constant can be
 derived from the current simplified method. Using numerical
simulations, Inagaki et al. (1995) and  
Roettiger {\it et al.} (1997) showed that the
Hubble constant measured through the SZ effect can be seriously 
affected by systematic effects, which include the assumption
of isothermality, cluster gas clumping, and asphericity.
The effects due to nonisothermality and density distribution, such as
gas clumping, can eventually be studied with upcoming high quality
X-ray imaging and spectral data from the {\it Chandra X-ray
Observatory}\footnote{http://asc.harvard.edu} (CXO)
and {\it X-ray Multiple Mirror
Mission}\footnote{http://astro.estec.esa.nl/XMM}. 
In addition to such expected improvements on the physical
state of the electron distribution responsible for the two scattering
and emission effects, one should consider the possibility that the
SZ/X-ray measurements are affected through cluster projection effects and
the intrinsic cluster shape distribution.

Using analytical methods, Cooray (1998)
and Sulkanen (1999) investigated projection effects on the Hubble
constant due to an assumption involving ellipsoidal shape for 
galaxy clusters. These studies led to the conclusion
that current measurements may be biased and that from a large sample of
clusters, it may be possible to obtain
 an unbiased estimate of the Hubble constant provided that
cluster ellipsoidal shapes can be identified 
accurately. Here, {\it large} depends on what was assumed in the
calculation; If the ellipticities of clusters follow the observed
distribution by Mohr et al. (1995), then a sample as
small as 25 clusters can, in  principle, provide a measurement of the
Hubble constant within few percent of the true value. 
The real scenario, however, can be much different
as the assumptions that have been made may be too simple. 

As an attempt to understand intrinsic
cluster shape distribution,  we used the available cluster data
to constrain the accuracy to which clusters can be described by simple
ellipsoidal models. Apart from previous work involving
cluster axial ratios measured through optical galaxy distributions 
(e.g., Ryden 1996), we note that no study has yet been performed on intrinsic
cluster shapes using gas distribution data, such as the X-ray isophotal
axial ratio distribution. 
Compared to optical galaxy isophotes, a study on cluster shapes
using X-ray data would be more appropriate as the gas distribution is
likely to be a better tracer of intrinsic cluster shapes.
Here, our primary goal is to quantify the nature of 
cluster shapes using X-ray observations reported in the literature.
We essentially follow the framework presented in Cooray (1998)
and describe intrinsic 
cluster shapes using axisymmetric models, mainly prolate (or
cigar-like) and oblate (pancake like) spheroidal distributions.
In Section 2, we briefly introduce the apparent cluster
shapes of axisymmetric galaxy clusters  and move on to discuss
intrinsic shapes. We also extended our discussion to
consider the possibility that clusters are triaxial ellipsoids with an
intrinsic distribution for axial ratios that follow a Gaussian form.
Given that the calculational methods to obtain intrinsic shapes given 
apparent or projected distributions are well known, especially for galaxies
and stellar systems such as globular clusters, 
we only present relevant details here. We refer the
interested readers to 
Merritt \& Tremblay (1994), Vio et al. (1994), Ryden (1992; 1996)
for further details and applications.
Given the wide and timely interest in using cluster SZ and X-ray 
data to derive cosmological parameters,
we follow well established procedures in these papers
to address what can be learnt of intrinsic shapes of clusters 
from current observational data.

\section{Galaxy Cluster Shapes}

\subsection{Apparent Shapes}
Given that there is a large amount of literature, including
textbooks (e.g., Binney \& Tremaine 1991), 
that  describe techniques to calculate the apparent
axial ratio distribution of projected bodies, mainly galaxies, 
we skip all the intermediate details and start by presenting
the expected distribution of apparent axial ratios for 
prolate and oblate spheroids.
In the case of a intrinsic prolate shape distribution, 
the apparent axial ratio distribution, $f(\eta)$, is:
\begin{equation}
f(\eta) = \frac{1}{\eta^2} \int_0^\eta
\frac{N_p(\gamma)\gamma^2\,d\gamma}{\left[(1-\gamma^2)(\eta^2-\gamma^2)\right]^{1/2}},
\end{equation}
while for the oblate distribution:
\begin{equation}
f(\eta) = \eta \int_0^\eta
\frac{N_o(\gamma)\,d\gamma}{\left[(1-\gamma^2)(\eta^2-\gamma^2)\right]^{1/2}}.
\end{equation}
In Eq.~1 \& 2, $N_p(\gamma)$ and $N_o(\gamma)$ represent,
respectively, the intrinsic axial ratio distribution when clusters are
assumed to be prolate and oblate.

In order to obtain the underlying distribution of apparent axial
ratios using a measured series of axial ratio values ($\eta$),
we use the nonparametric kernel estimator given by:
\begin{equation}
\hat{f}(\eta) = \frac{1}{Nh} \sum_{i=1}^{N}
K\left(\frac{\eta-\eta_i}{h}\right)
\end{equation}
where $K$ is the kernel function with
 kernel width $h$ (e.g., Merritt \& Tremblay 1994) and
$N$ is the total number of clusters.
For the present calculation, we use a smooth function to
describe the Kernel:
\begin{equation}
K(x)=\frac{1}{\sqrt{2\pi}} \exp\left(-\frac{x^2}{2}\right).
\end{equation}
In general, the kernel width is calculated by minimizing the mean integrated
square error (MISE), defined as the expectation value of the integral:
\begin{equation}
\int\left[\hat{f}(\eta)-f(\eta)\right]^2 d\eta.
\end{equation}
Such an estimation is problematic when $f(\eta)$ is
not known initially, and requires, usually, iterative schemes to
obtain the optimal $h$ value. Here, 
we take the approach presented
Vio et al. (1994) and used in Ryden (1996). Vio et al. (1994)
showed that a good approximation to kernel width for a wide range of
density distributions which are  reasonably smooth and not strongly skewed is:
\begin{equation}
h = 0.9AN^{-0.2}.
\end{equation}
Here, $A$ is chosen such that it is the smaller of either the standard
deviation of the sample or
the interquartile range of the sample divided by 1.34. 
Accordingly, this approximation is expected
to usually produce an estimate
within 10\% of the distribution when $h$ is calculated by minimizing
MISE.

Since $\eta$ is limited by definition to the range between 0 and 1,
we use the so-called reflective boundary conditions at $\eta=0$ and
$\eta=1$ (e.g., Silverman 1986). This is done by replacing the
Gaussian kernel $K$ above with the kernel (Ryden 1996):
\begin{eqnarray}
K^{\prime}(\eta,\eta_i,h) & =
K\left(\frac{\eta-\eta_i}{h}\right)+K\left(\frac{\eta+\eta_i}{h}\right)
\\ \nonumber
 & +K\left(\frac{2-\eta-\eta_i}{h}\right),
\end{eqnarray}
such that the Gaussian tails that extended less than 0 and greater than
1 are folded back into the interval between 0 and 1, with 0 and 1
inclusive. Such reflective boundary conditions ensure that the
proper normalization is uphold:
\begin{equation}
\int_0^1\hat{f}(\eta)\,d\eta = 1
\end{equation}
as long as $h << 1$. However, these reflective boundary conditions
forces the estimated distribution to have zero derivatives at
the two boundaries. Such artificial modifications
may be problematic when interpreting the observed distribution
near boundaries of 0 and 1; One 
should be cautious on the accuracy of the
estimated distribution and the inverted profiles near such values.

\begin{figure}
\psfig{file=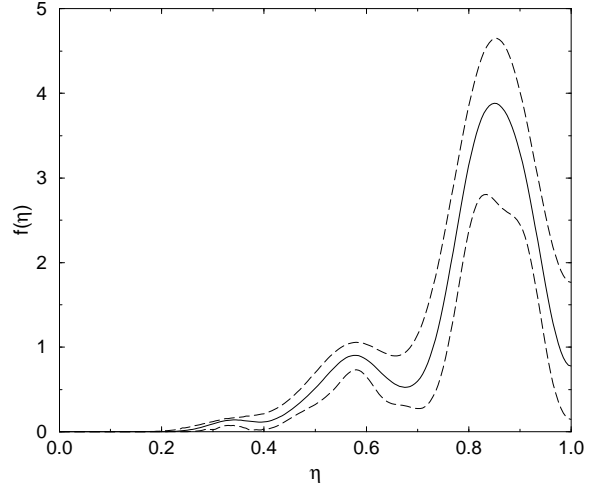,width=2.6in,angle=-90}
\caption{Nonparametric kernel estimate of the distribution of
apparent axial ratios for a sample of galaxy clusters ({\it solid
line}, using X-ray isophotal data from Mohr et al. (1995). 
The long-dashed lines show the
90\% confidence range on the estimate using a bootstrap monte-carlo
calculation of the observed distribution.}
\end{figure}

\subsection{Intrinsic Shapes}
In order to obtain the intrinsic distribution, one can easily invert
Eqs.~1 \& 3, respectively. Such an inversion can now be carried out
directly as we now have an estimator for the underlying distribution of
apparent axial ratios. 

If clusters are all randomly oriented ellipsoids following a strict 
oblate distribution, then the estimate distribution for the intrinsic
axis ratio, $\hat{N}_0(\gamma)$ is given by the relation:
\begin{equation}
\hat{N}_o(\gamma) = \frac{2\gamma\sqrt{1-\gamma^2}}{\pi} \int_0^\gamma
\frac{d}{d\eta}\left(\frac{\hat{f}(\eta)}{\eta}\right)\frac{d\eta}{\sqrt{\gamma^2-\eta^2}}.
\end{equation}
However, if clusters are assumed to randomly oriented ellipsoids
following a prolate distribution, then the intrinsic distribution is:
\begin{equation}
\hat{N}_p(\gamma) = \frac{2\sqrt{1-\gamma^2}}{\gamma \pi} \int_0^\gamma
\frac{d}{d\eta}\left(\eta^2\hat{f}(\eta)\right)\frac{d\eta}{\sqrt{\gamma^2-\eta^2}}.
\end{equation}
Other than such a direct inversion, 
various other iterative (e.g., Lucy's method; Lucy 1974) 
techniques can also be used to obtain the intrinsic
distribution. However,
for the purpose of this calculation, we use the direct inversion using
above integrals.

To be physically meaningful, $\hat{N}_o$ and $\hat{N}_p$ should be
nonnegative over the entire range of $\gamma$ values from 0 to 1.
Since we directly compute $\hat{N}_o$ and $\hat{N}_p$ without
making any restrictions on the values it can take between $\gamma$ of 0
and 1, our approach allows us to test the null hypothesis that all objects are
either oblate or prolate. However, we note that certain iterative schemes
available in the literature, which can be utilized
for an inversion of the observed axial ratio distribution,
do not necessarily make such a test possible as such schemes 
impose a priori constraint  that $\hat{N}_o$, or $\hat{N}_p$, is 
positive for all values between 0 and 1.

To impose a reasonably accurate constraint that objects cannot be
either prolate or oblate, we conduct
a monte carlo study of the observed data by using a bootstrap
resampling procedure; From the original data set of $\eta_i$ values
fom Mohr et al. (1995) sample,
we draw, with replacement, a new set of axial ratios that represent
the same data set. Here, we now consider the uncertainties associated
with Mohr et al. (1995) axial ratio measurements and allow these
bootstrap samples to take axial ratio values which are within $\pm$ 1
$\sigma$  of the measurement error range. These points are then used to create
a new bootstrap estimate for $\hat{f}$ (Fig.~1), which is inverted to compute
estimates for $\hat{N}_o$ and $\hat{N}_p$. We create a substantial
number of such bootstrap datasets to place robust confidence intervals
on the original dataset. At each value of $\gamma$,
confidence intervals are placed on either
$\hat{N}_o$ or $\hat{N}_p$ by finding values of $\hat{N}_o$ or
$\hat{N}_p$ such that the bootstrap estimates lie above some 
confidence limit. If this confidence limit drops below 
zero for any value of $\gamma$ between 0 and 1, the hypothesis that all objects
are oblate, or prolate, can be rejected (see, Ryden 1996).  
For the purpose of this paper, we use $10^4$ bootstrap
resamplings, in order to have sufficiently accurate measurements of
the underlying distribution function to impose confidence levels
at which either the prolate or the oblate hypothesis is rejected. 
This approach is essentially similar to what Ryden (1996) has utilized
to constrain the intrinsic shapes of various sources, such as globular
clusters and elliptical galaxies.

\begin{figure}
\psfig{file=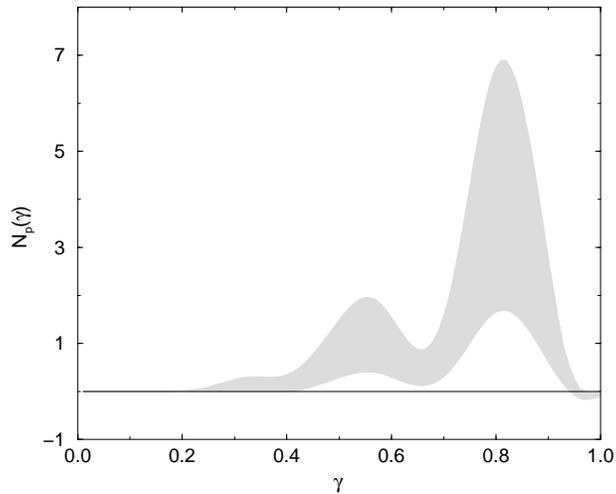,width=2.6in,angle=-90}
\caption{Distribution of intrinsic axial ratios, assuming that
clusters are prolate. The shaded region represent the 90\% confidence
band found by bootstrap resampling of the Mohr et al. (1995) cluster sample.}
\end{figure}

\begin{figure}
\psfig{file=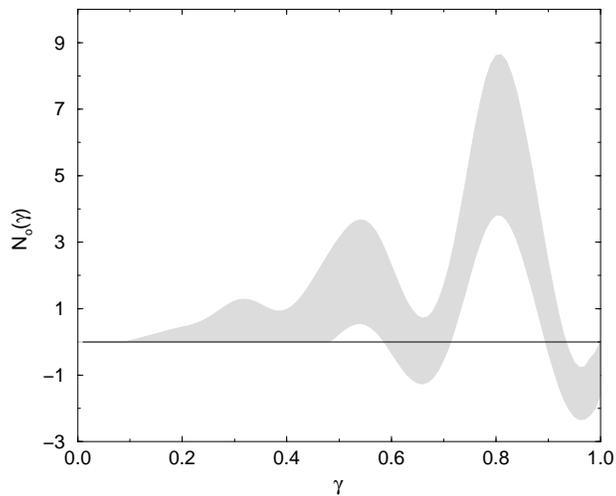,width=2.6in,angle=-90}
\caption{Distribution of intrinsic axial ratios, assuming that
clusters are oblate (same as Fig.~2).}
\end{figure}

In order to obtain constraints on the intrinsic cluster shapes, we use
the Mohr et al. (1995) cluster sample. Here, the authors studied 65 nearby
clusters and presented apparent axial ratios of these clusters using
X-ray isophotal data. This is the largest such study available in the
literature, while other studies, involving a less number of clusters,
 essentially contains more or less the
same clusters as the Mohr et al. (1995) sample.
Another advantage of the Mohr et al. (1995) cluster sample is that it
is X-ray flux limited and clusters were not selected based on the
X-ray
surface brightness. The original sample in Mohr et al. (1995) was
defined
by Edge et al. (1990) based on observations by {\it HEAO-1} and {\it
Ariel-V}
surveys combined with {\it Einstein Observatory} imaging observations.
Such a flux-limited complete, or near-complete, sample, instead of a
surface brightness selected sample, has the advantage that clusters
are
not likely to be biased in their selection. Such selection effects,
say due to elongated nature by enhancing the surface brightness,
would be problemtic both for the current study on the
intrinsic shapes of clusters as well cosmological studies using
clusters based on the X-ray luminosity and temperature function.
For the purpose of this paper, we assume that clusters in the Mohr et
al. (1995) sample has been selected in an unbiased manner when their
intrinsic shapes are considered (see, also, Edge et al. 1990).
  
We use tabulated axial ratio measurements in Table~3 of Mohr et
al. (1995), which contains measurements for 58 clusters, to obtain a
nonparametric estimates for the underlying distribution. These were
then inverted to obtain intrinsic axial ratio distributions,
assuming prolate and oblate shapes for clusters.
In Figs.~2 \& 3, we show our results; the shaded region represent the
90\% confidence limits from bootstrap resampling technique.
If we assume that all clusters are prolate, the observed distribution
is consistent with such an assumption; except when $\gamma \sim 1$,
the distribution is always positive. However, if we assume that all
clusters are oblate, then the resulting intrinsic distribution is
inconsistent with such an assumption at the $\sim$ 98\% 
confidence. Returning to previous works, we find that such a conclusion is
consistent with constraints on intrinsic cluster shapes using optical
data. In Ryden (1996), for various optically selected samples,
randomly oriented oblate hypothesis was rejected at a higher
confidence level than the randomly oriented prolate hypothesis. 
However, we note an alternative possibility that clusters
are in fact triaxial ellipsoids. Another possibility is that our
assumption that clusters are randomly oriented ellipsoids  may be
incorrect; clusters can still be oblate ellipsoids, however,
they should be oriented in preferred directions than random directions.
Since we do not have additional information on such scenario,
we may be left with the possibility that clusters are either
randomly oriented prolate or randomly oriented triaxial ellipsoids. 

\subsection{Clusters as Triaxial Ellipsoids}

In order to test the possibility that galaxy clusters are triaxial
ellipsoids viewed from random angles, we now consider random projections
of such objects. It has been shown in Stark (1977; also, Binney
1985) that triaxial ellipsoids  project into ellipses when viewed at
random angles. Assuming a viewing angle of $(\theta,\phi)$, in a
standard polar coordinate system with $z$-axis acting as the pole,
the axial ratio of such an ellipse can be written as (Binney 1985;
Ryden 1992):
\begin{equation}
q(\beta,\gamma,\theta,\phi) = 
\left[\frac{A+C-\sqrt{(A-C)^2+B}}{A+C+\sqrt{(A-C)^2+B}}\right]^{1/2}
\end{equation}
where,
\begin{eqnarray}
A = &
\left(\cos^2\phi+\beta^2\sin^2\phi\right)cos^2\theta+\gamma^2\sin^2\theta,
\nonumber \\
B = & 4 \cos^2\theta\sin^2\phi\cos^2\phi \left(1-\beta^2\right)^2 \\
C = & \sin^2\phi + \beta^2 \cos^2\phi \nonumber \\
\end{eqnarray}
and $\beta$ and $\gamma$ are the intrinsic axis ratios of the ellipsoid.
Following Ryden (1992), where a similar calculation was applied to
elliptical galaxies to address their intrinsic shape distribution,
 we test the possibility that clusters are intrinsically triaxial
ellipsoids with axis ratios of ellipsoids distributed
according to a Gaussian distribution:
\begin{equation}
f(\beta,\gamma) \propto \exp\left[-\frac{(\beta-\beta_0)^2 + (\gamma -
\gamma_0)^2}{2\sigma_0^2}\right],
\end{equation}
and the constraint $1 \geq \beta \geq \gamma \geq 0$.
Here, $\beta_0$, $\gamma_0$ and $\sigma_0$ describe the intrinsic
Gaussian distribution and whose parameters can be constrained by a
comparison of the observed axial ratios given by Eq.~11.
For a set of $\beta_0$, $\gamma_0$ and $\sigma_0$ values, 
we randomly generate
$(\beta,\gamma)$ values that follow the above Gaussian distribution
and the associated constraint. We them view each pair of
$(\beta,\gamma)$ values with randomly chosen set of viewing angles
$(\theta,\phi)$. Following this procedure, we randomly generate $\sim$
10$^5$ $q$ values for which we apply the non-parametric kernel
estimator to obtain the underlying distribution.
Using $\chi^2$ statistic, we compare this underlying distribution to
the observed distribution and its error from the Mohr et al. (1995)
dataset. Finally, we repeat this procedure for different values of the
basic parameters that define the Gaussian distribution.

In Fig~4, we show constraints obtained on the intrinsic shape
parameter distribution by comparing to present observations. Here, we
show the 99\%, 95.4\% and 99.99\% confidences on $\beta_0$ and
$\gamma_0$ for several values of $\sigma_0$. As shown, the observed
distribution of axial ratios are consistent when $\beta_0$ is at the
high end, while $\gamma_0$ varies from low values to high values as
$\sigma_0$ is increased. For low $\sigma_0$ values, the observations
are more consistent with the possibility that clusters are oblate
($\beta_0=1$) rather than prolate ($\beta_0=\gamma_0$). However, as
$\sigma_0$ is increased  the observed distribution becomes more
consistent with the possibility that clusters are intrinsically
prolate. Still, we note that there is
a large range of possibilities where the observations are consistent
with values for $\beta_0$ and $\gamma_0$ which are neither consistent 
with the prolate nor the oblate hypothesis. For the parameter space 
considered here, the best fit
model has same $\beta_0$ and $\gamma_0$ values of 0.92 and
$\sigma_0=0.21$.  The reduced $\chi^2$ value of this model and data is
1.1. In general, when $\sigma_0 > 0.1$, statistically acceptable
fits are found when $\beta_0$ is close to $\gamma_0$, suggesting that
current cluster data are more consistent with an
intrinsically prolate distribution. 

\begin{figure}
\psfig{file=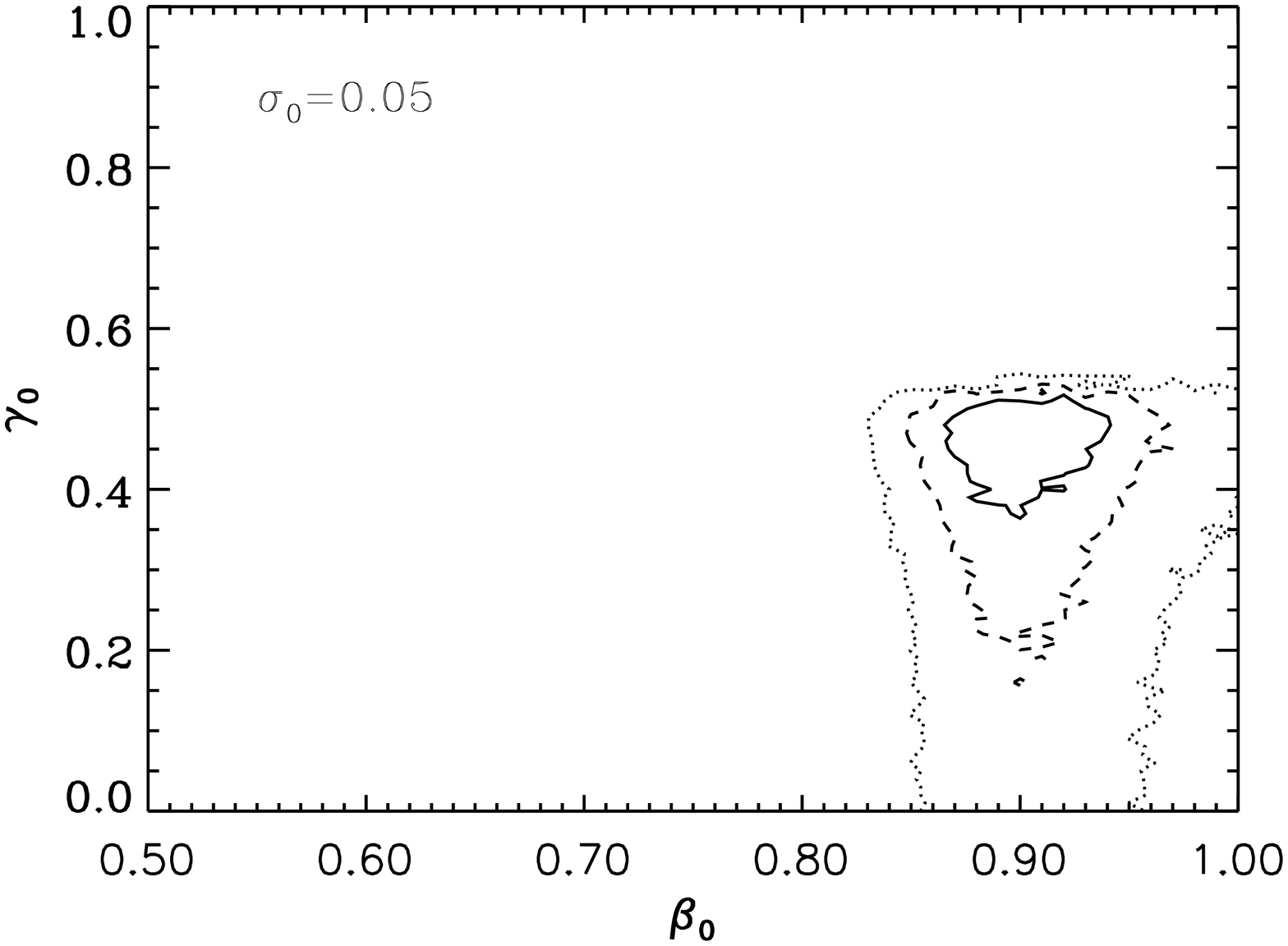,width=2.6in,angle=90}
\psfig{file=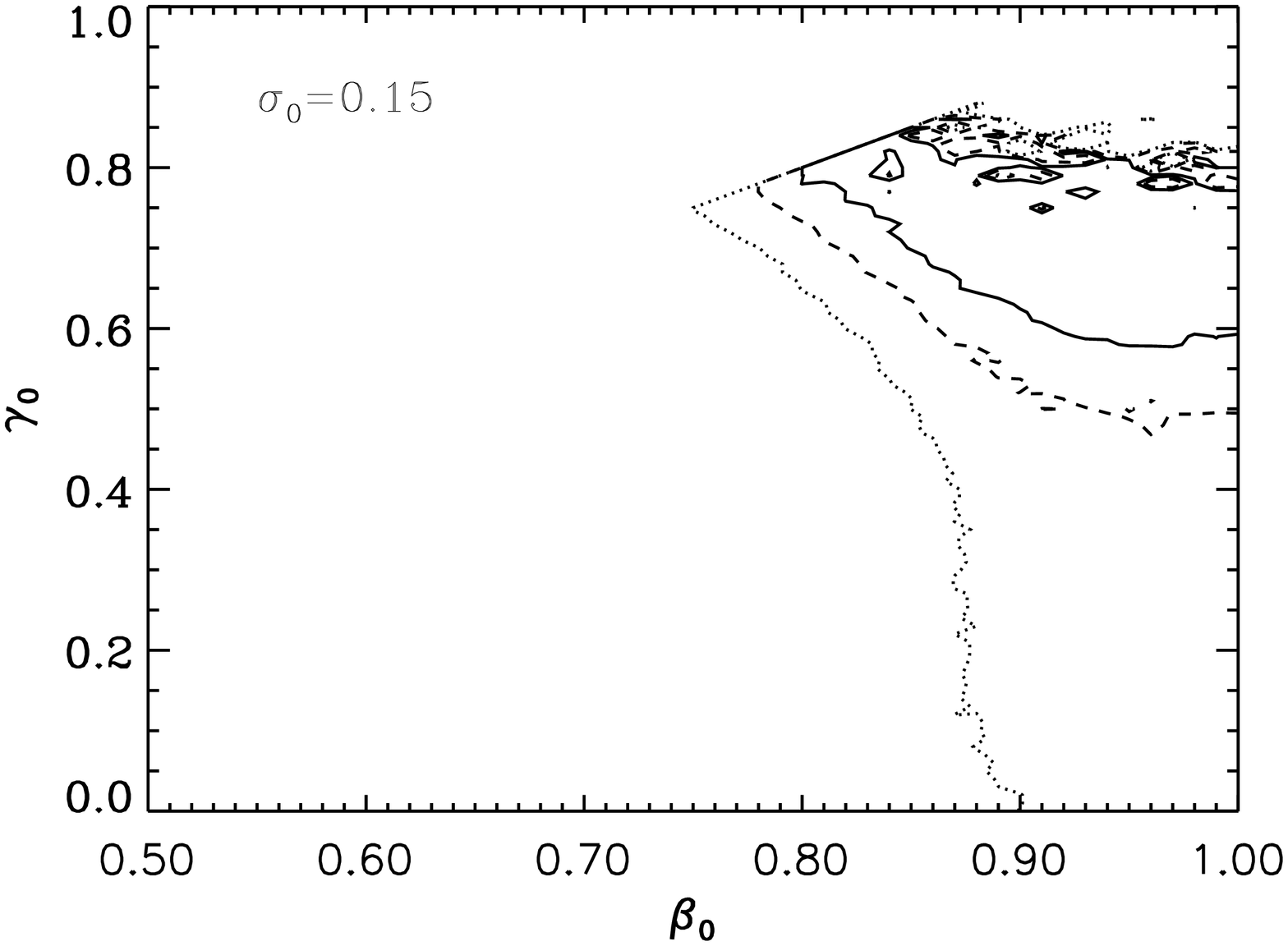,width=2.6in,angle=90}
\psfig{file=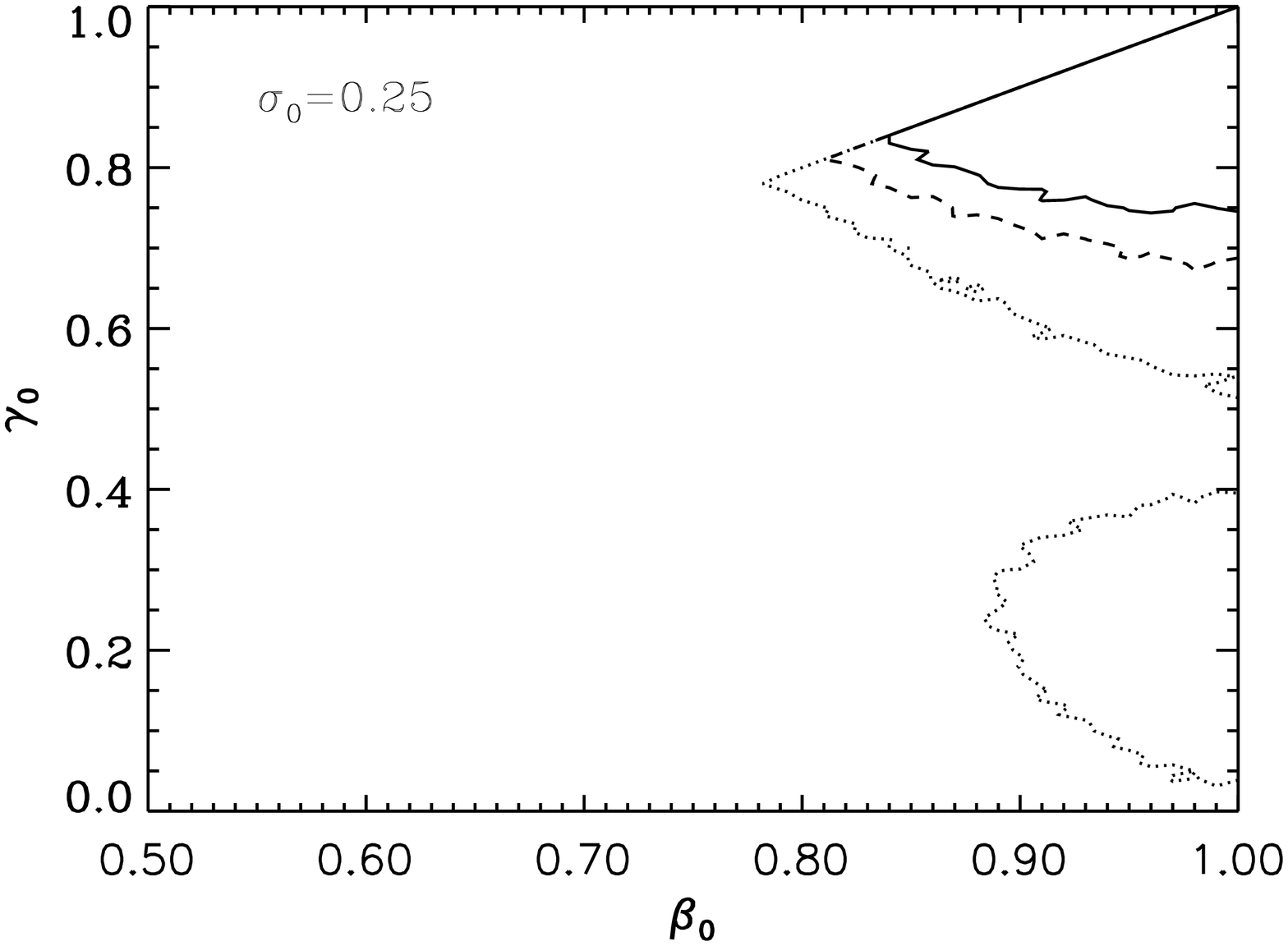,width=2.6in,angle=90}
\caption{Confidence limits on $\beta_0$, $\gamma_0$ and $\sigma_0$
parameter space. Shown here are the $\beta_0$ and $\gamma_0$
confidences for three values of $\sigma_0$ (0.05, 0.15 \& 0.25). 
The contours are at 95.4\%, 99\% and 99.99\% respectively.} 
\end{figure}

\section{Discussion \& Summary}

Using the Mohr et al. (1995) cluster sample, we rule out that clusters
are intrinsically axisymmetrical oblate ellipsoids at the 98\%
confidence level. As the Mohr et al. (1995) cluster sample is a
flux limited  sample rather than a surface brightness selected sample,
we can consider 
 such a sample as a fair representation of clusters in the Universe.
Mohr et al. (1995)  cluster sample also describes 
clusters which are now observed
both for the SZ effect and the X-ray emission and are used for the 
determination of the SZ/X-ray Hubble constant. 
Thus, conclusions based on the Mohr et al. (1995) sample should be
valid for what one can expect from current attempts to determine
cosmological parameters using SZ and X-ray data of galaxy clusters.
We have assumed that cluster X-ray isophotes represent the true shape
of galaxy clusters. It may be likely that cluster X-ray isophotes are
flattened compared to the intrinsic cluster shapes, and by ignoring
this possibility, we may have introduced a systematic bias in this
study. However, we note that such bias, if exists, is likely to be
small and that compared to other cluster data available to conduct a
study on intrinsic cluster shapes, X-ray isophotal axial ratios
allow a strong possibility to obtain reliable
conclusions on cluster shapes. Also, we note that any correction to
the measured Hubble constant due to asphericity is likely to be
based on the shape of X-ray isophotes, which is also expected to
be similar to SZ isophotes as both essentially measure the same
distribution. Therefore, the use of X-ray isophotes 
to constrain intrinsic shape distribution should be accurate and
valid, when considering the cosmological applications.

Our study shows that clusters are more likely to be prolate rather
than oblate ellipsoids, however, we cannot  rule out the possibility
that clusters are intrinsically triaxial.
Considering our previous discussions in Cooray (1998) related to
cluster projection effects on the SZ/X-ray Hubble constant, 
intrinsic prolate distributions allow a less biased
 determination of the Hubble constant, while an intrinsic oblate 
distribution results in a mean value for the Hubble constant which can
be biased as large as $\sim$ 10\% from the true value.
In Cooray (1998), we only considered the projection
effect arising from the unknown inclination angle of galaxy clusters
by averaging over a uniform distribution in inclination angles,
while only considering a mean value for the axial ratio of clusters
from Mohr et al. (1995).
Given that we have now determined the intrinsic distribution of
axial ratios,  we can now extend our calculations presented in Cooray
(1998) to also consider intrinsic axial ratio distribution.
Here, we assume that SZ and X-ray shape parameters coincide, however,
this is only true if clusters are triaxial ellipsoids. If the true shape of
clusters were to be more complicated, then a detailed analysis would
be necessary to obtain the individual shape parameters associated with
SZ and X-ray data and to determine the Hubble constant.

Assuming a simple scenario in which clusters are triaxial ellipsoids,
for a cluster sample of 25 clusters randomly drawn from the
intrinsic prolate and oblate distributions, 
we find that the oblate assumption and its 
distribution results in a biased measurement of the Hubble constant
by $\sim$ 8\%, while  for a prolate distribution, the resulting mean
value for the Hubble constant is unbiased, or within $\sim$ 3\%. 
For both prolate and oblate distributions,  
the width of the resulting distribution of Hubble constant values
agree with each other. These estimates both over and underestimates
 such that true value is within the range. These calculations 
and ones presented elsewhere (e.g., Sulkanen 1999) suggest 
 that the measurement of the Hubble constant based on galaxy clusters 
is not fundamentally biased by cluster projection effects and the
shape  distribution. 
Therefore, it is likely that a reliable measurement of the Hubble
constant will soon be possible with galaxy clusters using SZ and X-ray
data, however, such a calculation would still require that we improve
our knowledge on cluster physical properties such as isothermality and
gas clumping.

\section*{Acknowledgments}

I would like to acknowledge useful discussions with 
Scott Dodelson on inversion techniques and Joe Mohr on cluster
projections.

\end{document}